\documentclass[reprint,aps,pra,prb]{revtex4-1}

\usepackage{graphicx}
\usepackage{dcolumn}
\usepackage{bm}
\usepackage{amsmath}
\usepackage{amsthm}
\usepackage{amssymb}
\usepackage{float}
\usepackage{hyperref}
\usepackage{color}
\usepackage{epstopdf}
\usepackage{cleveref}
\usepackage[svgnames]{xcolor}
\hypersetup{hidelinks,colorlinks=true,allcolors=DarkBlue}

\begin{document}

\preprint{APS/123-QED}

\title{Two-dimensional Bose gas of tilted dipoles: roton instability and condensate depletion}
\author{A.\,K.\,Fedorov$^{1}$}
\author{I.\,L.\,Kurbakov$^{2}$}
\author{Y.\,E.\,Shchadilova$^{1}$}
\author{Yu.\,E.\,Lozovik$^{2,3,4,*}$}
\affiliation
{
\mbox{$^{1}$Russian Quantum Center, Skolkovo, Moscow 143025, Russia}
\mbox{$^{2}$Institute of Spectroscopy, Russian Academy of Sciences, Troitsk, Moscow Region 142190, Russia}
\mbox{$^{3}$National Research Nuclear University ``MEPhI'', Moscow 115409, Russia}
\mbox{$^{4}$MIEM at National Research University HSE, Moscow 109028, Russia}
}
\date{\today}

\begin{abstract}
We predict the effect of the roton instability for a two-dimensional {\it weakly} interacting gas of tilted dipoles in a single homogeneous quantum layer.
Being typical for strongly correlated systems, the roton phenomena appear to occur in a weakly interacting gas.
It is important that in contrast to a system of normal to wide layer dipoles, 
breaking of the rotational symmetry for a system of tilted dipoles leads to the convergence of the condensate depletion even up to the threshold of the roton instability, 
with mean-field approach being valid. 
Predicted effects can be observed in a wide class of dipolar systems.
We suggest observing predicted phenomena for systems of ultracold atoms and polar molecules in optical lattices, and estimate optimal experimental parameters. 

\begin{description}
\item[PACS numbers]
03.75.Kk, 03.75.Nt, 05.30.Jp
\end{description}
\end{abstract}
                              
\maketitle

\section{Introduction}
Bosonic systems with the dipole-dipole interaction are highly promising for observation of novel quantum phases and many-body phenomena \cite{LM,Santos0,revs}.
Due to significant experimental progress, several realizations of these systems have been studied: 
ultracold atoms \cite{manybodyrevs} with large magnetic dipole moment ({\it e.g.}, chromium, dysprosium and erbium), 
for which Bose-Einstein condensation (BEC) has been recently demonstrated \cite{prl094160401,prl107190401};
ultracold clouds of ground-state diatomic polar molecules \cite{polmol} in electric fields \cite{pra075012704,np0007000502,pt0064000027,KRb,OH,prl103183201} ({\it e.g.},  KRb and RbOH);
Rydberg atoms in electric fields \cite{Rydberg};
and excitons with spatially separated electrons and holes in semiconductor layers \cite{LY, Butov,FKL}.
These systems are well controllable via external fields. 
In particular, $s$-wave scattering length can be controlled via Feshbach resonances \cite{na0448000672, Chin}.

The anisotropy and the region of attraction of the dipole-dipole interaction provide a set of interesting collective phenomena.
In the limit of strongly correlated (classical) system of in-plane dipoles, the ground state of the system has the chain structure \cite{LM}; the 3D system of parallel dipoles has the chain structure as well \cite{L}. 
In Ref. \cite{Santos}, the roton-maxon spectrum has been predicted for normal to wide layer (pancake) dipoles.
The key feature of dipolar BEC is the character of the excitation spectrum \cite{Spec} similar to that in superfluid $^4$He \cite{He}.
 
Interesting structural properties emerge close to the threshold of the instability. 
The stability criterion is the non-negativity of square of the Bogoliubov spectrum,
\begin{equation}\label{epB}
	\varepsilon_{p}^2=\frac{p^2}{2m}\left(\frac{p^2}{2m}+2n_0{U}(p)\right)\ge0,
\end{equation}
where $p$ is the momentum, 
$m$ is the mass of dipoles, 
$n_0$ is the condensate density, 
$U(p)$ is the Fourier transform of an interaction pseudopotential. 
On the one hand, the stability problem has inspired great progress in investigation of superfluidity \cite{superfluidity}, density waves \cite{DWs}, phonon collapse \cite{Santos0, phc3Dd}, vortices \cite{vortices}, behavior of the system in optical lattices \cite{pra084053601}, traps \cite{trap}, and presence of disorder \cite{disorder}.
Monte Carlo simulations have predicted that a 2D gas of dipoles exhibits a quantum phase transition to a triangle crystal phase at zero temperature \cite{QMC}. 
Special attention \cite{SSclust,SShom,SSlatt} has been paid to a supersolid phase \cite{Andreev}.

\begin{figure}[t]
\begin{centering}
\includegraphics[width=1\columnwidth]{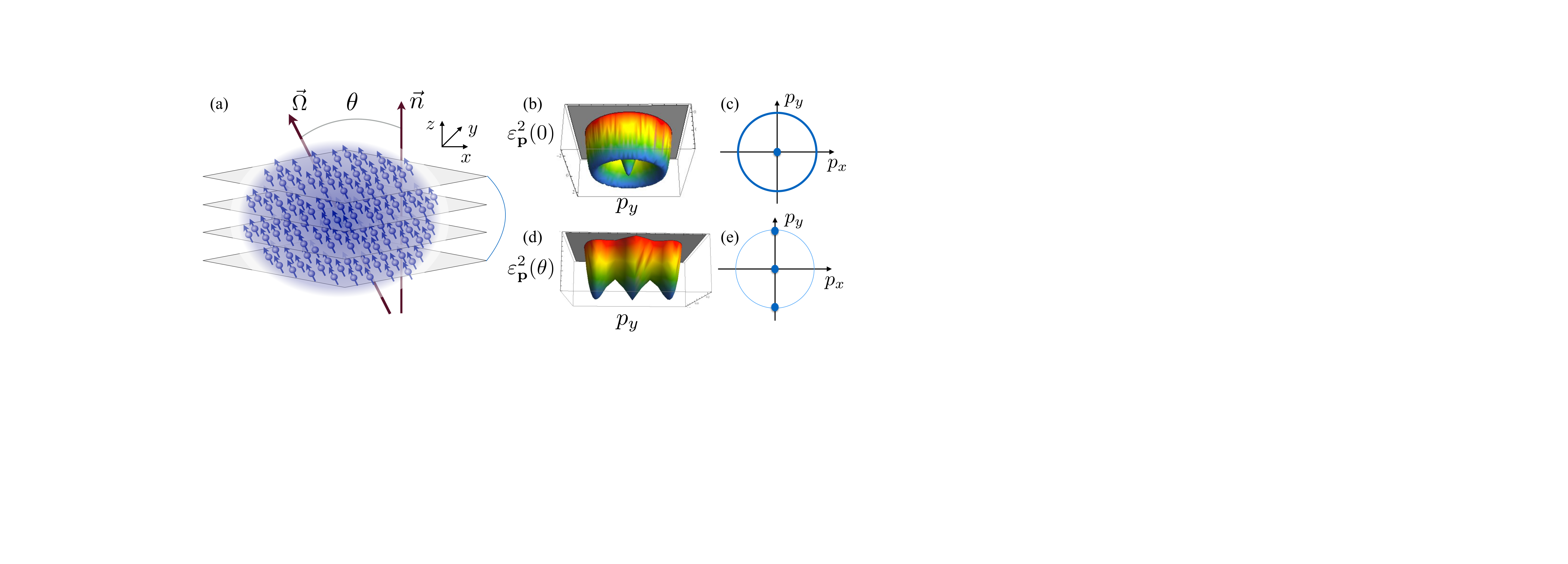}
\end{centering}
\vskip -2mm
\caption
{
In (a) BEC gas tilted in the $x{-}z$ plane dipoles in a 1D optical lattice with the harmonic trap in the $z$ direction. 
The angle $\theta$ to the layer is controllable by an external (electric or magnetic) field ${\vec\Omega}$. 
The square of spectrum $\varepsilon_{\bf p}^2$ [see Eq. (\ref{epB})] as a function of 2D momentum ${\bf p}=\{p_x,p_y\}$ at the threshold of the instability:
(b) surface plot of $\varepsilon_{\bf p}^2$ at $\theta=0$, 
(c) at $\theta=0$, $\varepsilon_{\bf p}^2$ reaches zero at the circumference $|{\bf p}|=p_{\rm r}$;
(d) surface plot of $\varepsilon_{\bf p}^2$ at fixed $\theta\ne0$;
(e) if $\theta\neq0$, $\varepsilon_{\bf p}^2$ reaches zero in two points $\{0,\pm{p}_{\rm r}\}$, arising at $x=0$ due to polarizing in the $x{-}z$ plane.
}
\end{figure}
 
\begin{figure*}[htbp]
\includegraphics[width=1.9\columnwidth]{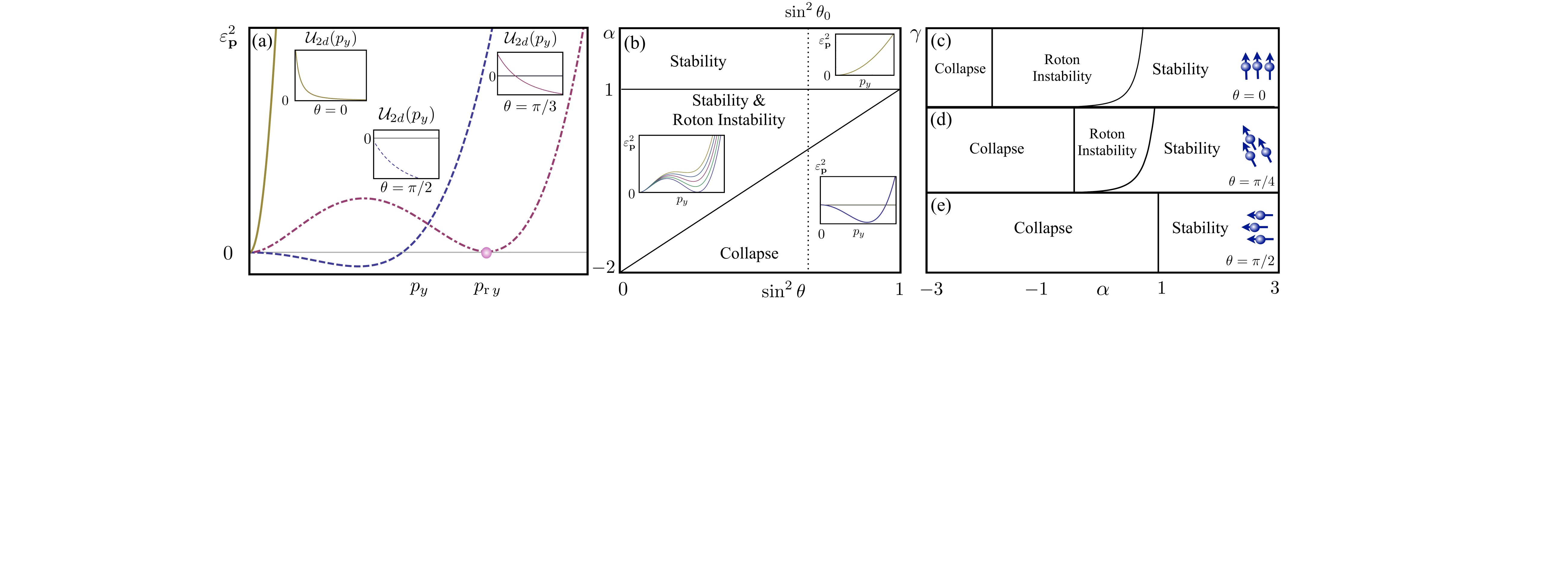}
\vskip -2mm
\caption
{
(a) The square of the Bogoliubov spectrum (main figure) and the effective interaction potentials (insets) for different tilt angle $\theta$:
a stable homogeneous gas at $\theta=0$ (solid); 
a phase with roton instability at $\theta=\pi/3$ (dot-dashed); 
a phase with long-wavelength collapse at $\theta=\pi/2$ (dashed).
Stability $\alpha-\theta-\gamma$ diagram (main figures) and typical spectrum branches (insets):
in (b) $\alpha-\theta$ projection, where at any $\gamma$ ({\it i.e.}, the diagram is independent on $\gamma$), the stable homogeneous and long-wavelength collapse phases
are shown, whereas in the solid upper triangular region (with the hypotenuse being $\sin^2\theta=(2+\alpha)/3$) the stable phase becomes roton unstable phase if $\gamma$ increases;
in (c)--(e) $\alpha-\gamma$ projection for (c) normal to the layer ($\theta=0$), (d) tilted ($\theta=\pi/4$), and (e) in-plane ($\theta=\pi/2$) dipoles;
the boundaries on the diagrams are defined via numerical solution of Eq. (\ref{rmri}).
Dotted line in (b) corresponds to the magic angle $\theta_0=\arccos(1/\sqrt{3})$, at which the contribution of the dipolar interaction in the plane reduces to zero [see Eq. (\ref{a})].
}
\label{p3}
\end{figure*}
 
On the other hand, the condensate depletion in the system diverges at the threshold of the roton instability \cite{pra073031602}. 
In other words, condensate disappears before the spectrum reaches zero even at zero temperature. 
Consequently, both the threshold of the roton instability and supersolid phase are unattainable \cite{Shlyapnikov2012}.

The above mentioned system of normal to wide layer dipoles is rotationally invariant. 
Therefore, the actual question is how this invariance affects system behavior and stability.
A simple example of a system with broken rotational symmetry is BEC of tilted dipoles \cite{prl106065301,pla376000480,pra084033625,vortices2, pra082043623,QMCtilt,Shlyapnikov2013,Fedorov}, 
where it is broken by the external (electric or magnetic) field. 
Great attention in these systems has been paid to anisotropy of superfluidity \cite{prl106065301}, sound velocity \cite{pla376000480}, correlators \cite{pra084033625}, vortices \cite{vortices2},
and the mean-field regime \cite{pra082043623}.
Monte Carlo studies have shown the difference in ground-states structures: 
normal to wide layer dipoles form a crystal \cite{QMC}, 
while tilting the dipoles has the effect of inducing striped structures \cite{QMCtilt}.
Related problems have been recently considered for dipolar fermions \cite{Shlyapnikov2013}.

In the present work, we consider dilute one-component BEC gas of 2D tilted in the $x{-}z$ plane at the angle $\theta$ dipolar bosons in a quantum layer at zero temperature $T=0$ [see Fig. 1(a)]. 
In fact, control for the angle $\theta$ is the way to tune the total scattering length $a$:
\begin{equation}\label{a}
	a=a_s+\left(3\cos^2\theta-1\right)a_d, \qquad a_d=\frac{md^2}{3\hbar^2},
\end{equation}
where $a_s$ is the $s$-wave scattering length and $a_d$ is the dipole-dipole scattering length ($d$ is the dipole moment). 
The system has three controllable dimensionless parameters:
(i) dipole tilting angle $\theta$ to normal (see Fig. 1), which is controllable by the polarizing (electric or magnetic) field;
(ii) ratio of scattering lengths $\alpha=a_s/a_d$, where $a_s$ is tunable by the Feshbach resonance and $a_d$ is controlled by an external field;
(iii) dimensionless density $\gamma=6\sqrt{2\pi}z_0a_dn_0$, which is controllable by changing the density (or tight-confinement oscillator length $z_0$ of the harmonic trap).
We consider negative values of the parameter $\alpha$ via negative values of $a_s$, whereas $a_d>0$ throughout.

Our paper is organized as follows. 
In Section \ref{secstability}, we demonstrate the roton-maxon character of the excitation spectrum of the system at finite $\theta$ as well as calculate the stability diagrams with respect to the controllable parameters of the system. 
In Section \ref{secdepletion}, we discuss the validity of the Bogoliubov approximation.
We show the convergence of the condensate depletion even up to the threshold of the roton instability, due to touching zero of the square of spectrum (\ref{epB}) only in two points.
We estimate experimental parameters for $^{164}$Dy atoms and for RbOH polar molecules in Section \ref{secexperiment}.
Finally, in Section \ref{secconclusion} we discuss and summarize our results.

\section{Stability problem}\label{secstability}
Let us consider a 3D gas of particles with both contact interaction, $g_{3d}\delta(\vec{\rm r}-\vec{\rm r}\,')$, and the dipole-dipole one,
$$
	V_{dd}(\vec{\rm r}\,,\theta)=\frac{d^2}{\vec{\rm r}\,^5}\left({\rm r}^2-3(x\sin\theta+z\cos\theta)^2\right).
$$

We use the following assumptions.
First, we consider the weak interaction limit $a_s,a_d\ll z_0$ \cite{prl084002551}.
Second, we imply the tight-confinement quantization, {\it i.e.}, $\hbar^2/mz_0^2$ is sufficiently larger that other energy scales of the problem, {\it e.g.}, the interaction energy 
(for details, see \cite{prl084002551,precise, Belyaev, LY2, scattering}).
At last, in the realization with 1D optical lattices [see Fig. 1(a)], we assume the independence of layers formed by the lattice potential, {\it i.e.}, 
we suppose that the interlayer tunneling is sufficiently small and the interlayer interaction of dipoles is totally screened. 

Under these assumptions, the Hamiltonian of the 3D system reads
\begin{eqnarray}\label{H3D}
	\mathcal{\hat H}_{3d}=\hat{H}_0+\hat{H}_{\rm int}, 
\end{eqnarray}
where
\begin{eqnarray}
	\hat{H}_0={\int}{d\vec{\rm r}\,\hat\Psi^+{(\vec{\rm r}\,)\left(-\frac{\hbar^2}{2m}\Delta_3+\mathcal{V}(\vec{\rm r}\,)-\mu_{3d}\right)\hat\Psi(\vec{\rm r}\,)}}, \nonumber  \\ 
	\hat{H}_{\rm int}=\frac{1}{2}{\int}{d\vec{\rm r}\,d\vec{\rm r}\,'\,\mathcal{U}(\vec{\rm r}-\vec{\rm r}\,',\theta)\hat\Psi^+(\vec{\rm r}\,)\hat\Psi^+(\vec{\rm r}\,')\hat\Psi(\vec{\rm r}\,')\hat\Psi(\vec{\rm r}\,)},  \nonumber
\end{eqnarray}
Here, 
$\hat\Psi(\vec{\rm r}\,)$ is the 3D field operator, 
$\Delta_3$ is the 3D Laplace operator, 
$\mu_{3d}$ is the 3D chemical potential, 
$\mathcal{V}(\vec{\rm r}\,)=V(\rho)+V_{\rm ti}(z)$ is the external potential,
$V({\rho})$ is the 2D confinement potential in thin layer plane,
$V_{\rm ti}(z)=(m\omega^2z^2)/2$ is the 1D confinement potential in the tight direction ($\omega$ is the oscillator frequency), 
and
$$
	\mathcal{U}(\vec{\rm r}-\vec{\rm r}\,',\theta)=V_{dd}(\vec{\rm r}-\vec{\rm r}\,',\theta)+g_{3d}\delta(\vec{\rm r}-\vec{\rm r}\,')
$$ 
is the interaction potential;
$\vec{\rm r}=\{{\rho},z\}$ and $\vec{\rm q}=\{{\bf p},p_z\}$ are 3D vectors, ${\rho}=\{x,y\}$ and ${\bf p}=\{p_x,p_y\}$ are 2D vectors.

In a sufficiently thin layer, the motion in the tight direction is frozen at the lowest energy state of the confining trap.
Thus, in the representation of the 3D field operator in the basis $\{\varphi_j^{\rm ti}(z)\}$ in the tight direction 
\begin{equation}\label{Psi2D}
	\hat\Psi(\vec{\rm r})\approx\varphi_0^{\rm ti}(z)\hat\Psi({\rho}), \qquad 
	\hat\Psi({\rho})=\int dz\varphi_0^{\rm ti*}(z)\hat\Psi(\vec{\rm r}),
\end{equation}
we take into account only the $j=0$ term.
Here, the field operator $\hat\Psi({\rho})$ is the effective 2D field operator, which satisfies the standard bosonic commutation relations.
The eigenfunctions $\varphi_j^{\rm ti}(z)$ and eigenenergies $\mathcal{E}_j^{\rm ti}$ are determined from the 1D Shr\"odinger equation:
$$
	\left(-\frac{\hbar^2}{2m}\frac{d^2}{dz^2}+V_{\rm ti}(z)\right)\varphi_j^{\rm ti}(z)=\mathcal{E}_j^{\rm ti}\varphi_j^{\rm ti}(z).
$$

By substituting (\ref{Psi2D}) in (\ref{H3D}), we find the effective Hamiltonian for the thin-layer motion,
\begin{equation}\label{H2D}
\begin{split}
	\mathcal{\hat H}_{2d}=\int{d{\rho}\,\hat\Psi^+({\rho})\left(-\frac{\hbar^2}{2m}\Delta_2-\mu+V({\rho})\right)\hat\Psi({\rho})}+ \\
	+\int{d{\rho}\,d{\rho}'\,\mathcal{U}_{2d}({\rho}-{\rho}',\theta)\hat\Psi^+({\rho})\hat\Psi^+({\rho}')\hat\Psi({\rho}')\hat\Psi({\rho})}
\end{split}
\end{equation}
with the effective 2D interaction potential,
\begin{equation}\label{U2D}	
	\mathcal{U}_{2d}({\rho}-{\rho}',\theta){=}\!\!\int\!\! dzdz'\,\mathcal{U}(\vec{\rm r}-\vec{\rm r}\,',\theta)|\varphi_0^{\rm ti}(z)\varphi_0^{\rm ti}(z')|^2.
\end{equation}
Here, $\Delta_2$ is the 2D Laplace operator, 
$\mu=\mu_{3d}-\mathcal{E}_0^{\rm ti}$ is the chemical potential, and
$$
	\mathcal{E}_0^{\rm ti}=\hbar\omega/2, \qquad \varphi_{\rm ti}(z)=\exp(-z^2/2z_0^2)/\sqrt{\sqrt{\pi}z_0}.
$$

Thus, we obtain the Fourier transform of the effective interaction potential of 2D dipoles (\ref{U2D}) in the Born approximation,
\begin{equation}\label{U2d}
	\mathcal{U}_{2d}({\bf p}, \theta)=g_s-g_d+U_h({\bf p})\sin^2\theta+U_v({\bf p})\cos^2\theta,
\end{equation}
where
\begin{eqnarray}
	U_h({\bf p})=\frac{2d^2}{\hbar}\int_{-\infty}^{+\infty}\frac{p_x^2dp_z}{p_x^2+p_y^2+p_z^2}\exp\left(-\frac{p_z^2z_0^2}{2\hbar^2}\right),  \nonumber \\ 
	U_v({\bf p})=\frac{2d^2}{\hbar}\int_{-\infty}^{+\infty}\frac{p_z^2dp_z}{p_x^2+p_y^2+p_z^2}\exp\left(-\frac{p_z^2z_0^2}{2\hbar^2}\right),  \nonumber
\end{eqnarray}
with coupling constants,
\begin{equation}
	g_s=\frac{2\sqrt{2\pi}\hbar^2a_s}{mz_0}=\frac{g_{3d}}{\sqrt{2\pi}z_0},\;
	g_d=\frac{2\sqrt{2\pi}\hbar^2a_d}{mz_0}=\frac{2\sqrt{2\pi}d^2}{3z_0}. \nonumber
\end{equation}

For the Bogoliubov spectrum square (\ref{epB}) with effective interaction potential (\ref{U2d}), we find the following regimes on the phase diagrams (see Fig. 2):

(i) The effective potential $\mathcal{U}_{2d}({\bf p})>0$ is positive for all momenta ${\bf p}$.
Hence, the square of the Bogoliubov spectrum $\varepsilon_{\bf p}^2>0$ is positive for all ${\bf p}$ as well.  
In this case, the homogeneous phase is stable at an arbitrary density.

(ii) At low momenta, the effective potential $\mathcal{U}_{2d}({\bf p})<0$ is negative.
Therefore, the square of the Bogoliubov spectrum $\varepsilon_{\bf p}^2$ drops below the zero point at momenta below some critical one.
This regime is known as phonon instability in respect to a long-wavelength collapse \cite{Santos0, phc3Dd}, which can appear in the system at an arbitrary density. 

(iii) At low momenta, the effective potential $\mathcal{U}_{2d}({\bf p})>0$ is positive, but $\mathcal{U}_{2d}({\bf p})<0$ is negative for a certain momentum range.
In this case, at the certain density, the square of the Bogoliubov spectrum  $\varepsilon_{\bf p}^2$ touches zero point of energy at nonzero momentum. 
This regime corresponds to the threshold of the roton instability. 

We find the boundaries for the roton instability on the diagrams (see Fig. 2) from the following equation
\begin{equation}\label{rmri}
	\varepsilon_{\bf p}^2(\theta)=\frac{d\varepsilon_{\bf p}^2(\theta)}{d{\bf p}}=0.
\end{equation}
Moreover, using these diagrams (see Fig. 2), one can compare results with important particular cases: systems of normal to wide layer and in-plane dipoles.

\section{Validity of the Bogoliubov approximation and condensate depletion}\label{secdepletion}

In our consideration, the main emphasis is accessibility of the threshold of the roton instability for tilted dipoles based on the challenging problem of validity of the Bogoliubov approximation.
In turn, the latter is related to two conditions: 
(i) absence of the divergence of the condensate depletion at the threshold of the roton instability and 
(ii) the negligibility of the loop diagrams. 

\subsection{Condensate depletion} 

Using the Bogoliubov transformation,
\begin{equation}\label{BT}
	\hat{a}_{\bf p}=u_{\bf p}\hat{b}_{\bf p}-v_{\bf p}\hat{b}^\dagger_{\bf -p}, \quad \hat{a}_{\bf p}=\int{\frac{d{\rho}}{\sqrt{V}}e^{-\frac{i}{\hbar}{\bf p}{\rho}}\hat{\Psi}(\rho)},
\end{equation} 
we obtain Hamiltonian (\ref{H2D}) without the external potential [{\it i.e.}, $V(\rho)=0$] in the diagonal form \cite{AGD}
\begin{equation}\label{H2Ddiag}
	\hat{\mathcal{H}}_{2d}=\sum_{{\bf p}\ne0}\varepsilon_{\bf p}(\theta)\,\hat{b}^\dagger_{\bf p}\hat{b}_{\bf p}+{\rm const}.
\end{equation} 
Here, $V$ is the volume of the quantization box, 
$\hat{b}_{\bf p}$ and $\hat{b}^{\dagger}_{\bf p}$ are the Bogoliubov excitation operators in the system, 
\begin{equation}\label{eBp2}
	\varepsilon_{\bf p}(\theta)=\sqrt{\frac{p^2}{2m}\left(\frac{p^2}{2m}+2n_0\,\mathcal{U}_{2d}({\bf p}, \theta)\right)}
\end{equation} 
is the Bogoliubov spectrum [see Eq. (\ref{epB})], and
the Bogoliubov $u,v$ functions have the form,
$$
	u^2_{\bf p}=\frac{\left(\varepsilon_{\bf p}+{p}^2/2m\right)^2}{2\varepsilon_{\bf p}\,{p}^2/m}, \quad 
	v^2_{\bf p}=\frac{\left(\varepsilon_{\bf p}-{p}^2/2m\right)^2}{2\varepsilon_{\bf p}\,{p}^2/m}.
$$

To obtain the occupation number of the noncondensate fraction $n_{\bf p}$, one needs to calculate the average $\langle{0}|\hat{a}^{\dagger}_{{\bf p}}\hat{a}_{{\bf p}}|{0}\rangle$,
where $|0\rangle$ is the ground state.
Taking into account that $|0\rangle$ is the vacuum of the Bogoliubov excitations [see Eq. (\ref{H2D})] and $\hat{a}_{{\bf p}}$ can be presented via the Bogoliubov excitation operators using transformation (\ref{BT}), we obtain 
\begin{equation}\label{np}
	n_{\bf p}=\langle{0}|\hat{a}^{\dagger}_{{\bf p}}\hat{a}_{{\bf p}}|{0}\rangle=|v_{\bf p}|^2=\frac{1}{2}\frac{\left(\varepsilon_{\bf p}-{p}^2/2m\right)^2}{\varepsilon_{\bf p}\,{p}^2/m}.
\end{equation} 
Using (\ref{eBp2}) and integrating (\ref{np}) on the momentum space, the equation for the condensate depletion follows: 
\begin{equation}\label{n-n0}
\begin{split}
	\frac{n-n_0}{n_0}=\frac{1}{2n_0}\int\frac{d{\bf p}}{(2\pi\hbar)^2}\frac{\left(\varepsilon_{\bf p}-{p}^2/2m\right)^2}{\varepsilon_{\bf p}\,{p}^2/m}= \\
	=\frac{1}{2n_0}\int\frac{d{\bf p}}{(2\pi\hbar)^2}\frac{p^2/2m+\mathcal{U}_{2d}({\bf p}, \theta)n_0-\varepsilon_{\bf p}}{\varepsilon_{\bf p}},
\end{split}
\end{equation}
where $n$ is the total density and $n_0$ is the condensate density.

\subsection{Normal to the layer dipoles ($\theta=0$)}

We start from consideration of the condensate depletion for normal to the layer dipoles.
In this case, square of spectrum (\ref{epB}) is an isotropic function of the momentum ${\bf p}$ [see Fig 1(b)].
Thus, close to the threshold of the roton instability ({\it i.e.}, at $p\approx p_{\rm r}$), we have the following approximation for (\ref{epB}):
\begin{eqnarray}\label{normaldepl}
	\varepsilon^2_{p}\big|_{{p}\approx\pm{p}_{\rm r}}
	\approx\frac{1}{2}\frac{d\varepsilon^2_{p}}{dp^2}({p}-p_{\rm r})^2. 
\end{eqnarray}

We substitute (\ref{normaldepl}) to condensate depletion (\ref{n-n0}), 
and find that the depletion for normal to the layer dipoles diverges at the threshold of the roton instability even at $T=0$ (see Fig. 3). 
This result was established for the first time in \cite{pra073031602}. 
Thus, one can conclude that for normal to layer dipoles the Bogoliubov approximation becomes inapplicable before the roton minimum reaches zero, 
{\it i.e.}, at sufficiently small nonzero $\Delta_{\rm r}>0$ roton gap \cite{Shlyapnikov2012}.

\subsection{Tilted to the layer dipoles ($\theta\ne0$)}

\begin{figure}[t]
\begin{centering}
\includegraphics[width=0.77\columnwidth]{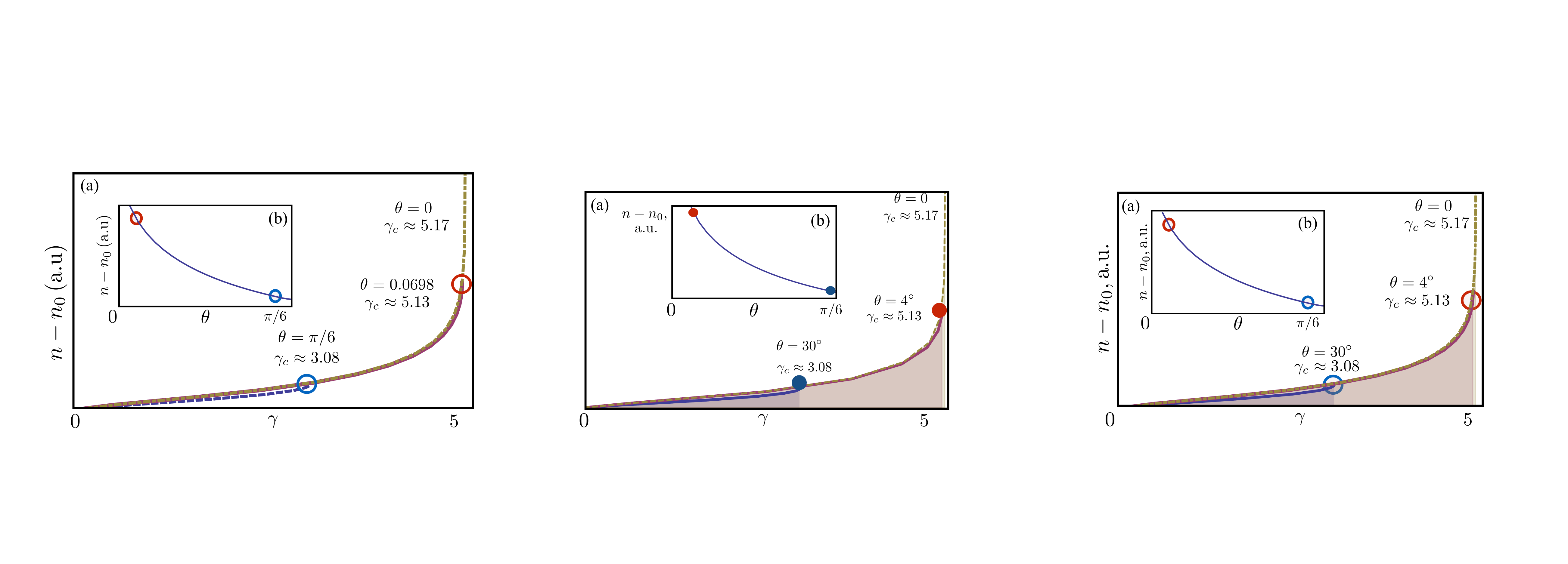}
\end{centering}
\caption{
Divergence of condensate depletion (\ref{n-n0}) for normal to wide layer dipoles (\ref{normaldepl}) and, in contrast, its convergence for tilted dipoles (\ref{tiltdepl}).
In (a) condensate depletion (\ref{n-n0}) is shown as a function of the dimensionless density $\gamma$: 
The depletion at $\theta=\pi/6$ with $\gamma_c\approx{3.08}$ (dashed line), circle (blue) is the value of depletion at the threshold of the instability; 
the depletion at $\theta\approx0.07$ with $\gamma_c\approx{5.13}$ (solid line), circle (red) is the value of depletion at the threshold of the instability;
the depletion at $\theta=0$ with $\gamma_c\approx{5.17}$ (dot-dashed line). 
In (b) the condensate depletion is shown as function of tilt angle $\theta$: first (red) circle on $\theta\approx0.07$; second (blue) circle on $\theta=\pi/6$.}
\end{figure} 

Due to the anisotropy of the square of spectrum (\ref{epB}) as a function of the momentum ${\bf p}$ for tilted dipoles [see Fig. 1(d)], close to the threshold of the instability, we obtain
\begin{eqnarray}\label{tiltdepl}
	\varepsilon^2_{{\bf p}}\big|_{{\bf p}
	\approx\pm{\bf p}_{\rm r}}&\approx&\frac{1}{2}\frac{\partial^2{\varepsilon^2_{{\bf p}}}}{\partial{p}_x^2}p_x^2+\frac{1}{2}\frac{\partial^2{\varepsilon^2_{{\bf p}}}}{\partial{p}_y^2}(p_y\mp p_{\rm r})^2= \nonumber \\
	&=&\mathcal{A}(\theta)p_x^2+\mathcal{B}(\theta)(p_y\mp p_{\rm r})^2,
\end{eqnarray}
where at ${\bf p}=\pm{\bf p}_{\rm r}$,
\begin{eqnarray}\label{A}
	\mathcal{A}(\theta)\equiv\frac{1}{2}\frac{\partial^2{\varepsilon^2_{{\bf p}}}}{\partial{p}_x^2}
	&=&\frac{p^2}{2m^2}+\frac{n_0p^2}{2m}\frac{\partial^2{\mathcal{U}_{2d}({\bf p}, \theta)}}{\partial{p_x^2}}, \\
	\mathcal{B}(\theta)\equiv\frac{1}{2}\frac{\partial^2{\varepsilon^2_{{\bf p}}}}{\partial{p}_y^2}
	&=&\frac{p^2}{2m^2}+\frac{n_0p^2}{2m}\frac{\partial^2{\mathcal{U}_{2d}({\bf p}, \theta)}}{\partial{p_y^2}},  \nonumber
\end{eqnarray}
with ${\partial^2{\varepsilon^2_{{\bf p}}}}/{\partial{p}_x\partial{p}_y}=0$.

By substituting (\ref{tiltdepl}) in (\ref{n-n0}), we obtain that in this case at the threshold of roton instability the condensate depletion (\ref{n-n0}) converges (see Fig. 3).  
Consequently, the depletion $n-n_0$ can be small enough if the interactions in the system are sufficiently weak. 
In this case, we obtain that $n_0\approx n$. 
Hence, for tilted dipoles, even at the threshold of the instability the Bogoliubov approximation is at least self-consistent.

The problem of negligibility of the loop diagrams in the weak interaction regime at $T=0$, when the roton minimum touches zero, is more complicated and it will be considered in another place.

Being complicated on the microscopic level, the problem of negligibility of the loop diagrams in the mesoscopic approach (see \cite{Popov,pra067053615}) is much more clear. 
Here, we based this on the following considerations. 
Let the parameters $\theta$, $\alpha$, and $\gamma$ of the problem be as follows that the roton gap $\Delta_{\rm r}$ is sufficiently small but differs from zero. 
Then, in both the macroscopic and weak interaction limits at $T=0$, all loop diagrams are vanishing. 
In this case, the Bogoliubov approximation (if it is self-consistent) is valid even in the macroscopic system. 

Let us consider a finite-size system, {\it e.g.}, a box of size $L_x{\times}L_y$ so that the following condition holds
$$
	 \mathcal{A}(\theta)(\hbar/L_x)^2+\mathcal{B}(\theta)(\hbar/L_y)^2 \gg \Delta_{\rm r}.
$$ 
In this case, both the condensate depletion and the loop diagrams are negligible, and, on the other hand, the system does not ``feels'' the presence of the roton gap $\Delta_{\rm r}$.
In this sense, in the mesoscopic formalism the Bogoliubov approximation is valid at the threshold of the roton instability. 
Moreover, it can be valid even though there is a macroscopic occupation in the region of the minimum. 

\section{Experimental realizations}\label{secexperiment}

We suggest experimental realizations of the roton minimum and the roton instability for dysprosium atoms and RbOH polar molecules.
Details of our estimations for the threshold of the roton instability are:

(i) Dysprosium atoms \cite{prl107190401}.
$m=164$ u, $z_0=150$ nm ($\hbar\omega=130$ nK, $\omega/2\pi=2.72$ kHz),
$\theta=72^{\circ}$, $a_d=7$ nm, $a_s=5.5$ nm, $a=0.5$ nm,
$n_0=2.15\times10^{10}$ cm$^{-2}$ ($\alpha=11/14$, $\gamma=17/5$),
$\mu=10.6$ nK, $n_0/n=197/200$.

(ii) Polar molecules RbOH \cite{pra075012704}.
$m=104$ u, $z_0=200$ nm ($\hbar\omega=116$ nK, $\omega/2\pi=2.42$ kHz),
$\theta=57.7^{\circ}$, $a_d=14$ nm, $a_s=5$ nm, $a=3$ nm,
$n_0=2.65\times10^{9}$ cm$^{-2}$ ($\alpha=5/14$, $\gamma=10/9$),
$\mu=9.3$ nK, $n_0/n=74/75$.

In recent experiments with ultracold molecules \cite{polmol}, difficulties are related to ultracold chemical reactions ({\it e.g.}, ${\rm KRb}{+}{\rm KRb}{=}{\rm K}_2{+}{\rm Rb}_2$).
Therefore, we expect that RbOH polar molecules are preferred for long-lived Bose gases.
In contrast to KRb molecules, ultracold RbOH molecules do not react as
${\rm RbOH}{+}{\rm RbOH}{=}{\rm Rb}_2{+}{\rm H}_2{\rm O_2}$ 
and 
${\rm RbOH}{+}{\rm RbOH}{=}{\rm Rb}_2{\rm O}{+}{\rm H_2}{\rm O}$, with the merging of two RbOH molecules into a dimer being suppressed.
Indeed, it is unlikely that the system merges into the excited dimer ${\rm RbOH}{+}{\rm RbOH}{=}{\rm Rb}_2{\rm H}_2{\rm O}_2^*$ 
because of huge (in comparison with typical energy scales of the problem) energy gaps in electronic degree (${\sim}10^4$ K), oscillation ($\sim$100 K), and rotational ($\sim$0.3 K) degrees of freedom \cite{pt0064000027}.
Hence, the chemical reaction of the merging of two molecules RbOH into the dimer should be accompanied by a photon emission. 
Therefore, reaction ${\rm RbOH}{+}{\rm RbOH}{=}{\rm Rb}_2{\rm H}_2{\rm O}_2{+}h\nu$ should be suppressed by the reaction barrier.
Thus, the lifetime of the system significantly increases.
Moreover, since all bonds of RbOH are saturated, at the merging of two molecules RbOH emitted photon has energy, which is much lower energy than at the merging of two radicals.
This fact suppresses merging into the dimer on the value of order $(h\nu)^3$ in comparison with common in literature OH \cite{OH} and NH \cite{prl103183201} radicals.

\section{Discussions and conclusion}\label{secconclusion}

In the broad sense, the Bogoliubov approximation as well as the mean-field approach can be valid in the quasi-2D system on the mesoscopic scales, {\it i.e.}, in the quasicondensate formalism (see \cite{Popov,pra067053615}). 
Therefore, when the coupling constants are sufficiently small, the roton minimum in the macroscopic system {\it should be closer to zero than the width of the crossover} of the roton instability at finite-size scales. 
Furthermore, the characteristic size $\delta p_{\rm r}$ for a domain of roton minimum should be smaller than the values $\hbar/L$ in the case of scales of order of $L$. 

Moreover, we admit there is a macroscopic occupation in the region $|{\bf p}-{\bf p}_{\rm r}|\lesssim\delta p_{\rm r}$ even though the true roton minimum is higher than zero. 
In this case, in the box with size $L\lesssim\hbar/\delta p_{\rm r}$, there exist two traveling waves with opposite momenta $\pm{\bf p}_{\rm r}$, which form a standing wave. 
The latter leads to existence of {\it density waves} in the box, because the noncondensate fraction in the regime of the sufficiently weak interaction is small. 
The obtained result on the convergence of condensate depletion (\ref{n-n0}) for titled dipoles supports this prediction.
Density waves in the box imply {\it local} density waves in a macroscopic system of tilted dipoles, {\it i.e.}, both diagonal and off-diagonal {\it short-range} orders. 
These orders are totally controlled by the external fields: 
The wave period $\lambda=2\pi\hbar/p_{\rm r}$ is given by the parameters $\alpha$ and $\theta$, the wave direction is determined by an orientation of the polarizing field, 
and the number of waves is controlled by the interaction weakness [{\it i.e.}, the quantity $(n-n_0)/n_0$]. 
At the same time, global density waves can be absent in the system.

Experimentally, the local density waves can be observed 
(i) in the weakly interacting system of size $L<\hbar/\delta p_{\rm r}$ or 
(ii) in measurements of short-range order in one-body density matrix or pair-correlation function \cite{prl091010406}.

In contrast to our case, for normal to layer dipoles, the Bogoliubov approximation is not universally self-consistent. 
This results from the divergence of the condensate depletion at the threshold of the roton instability \cite{pra073031602}. 
Besides, the above mesoscopic arguments do not justify the Bogoliubov approach even in the weak interaction regime. 
It is in agreement with \cite{Shlyapnikov2012}.

To summarize, we have considered the stability problem for BEC gas of 2D tilted dipoles in the quantum layer. 
We have obtained stability diagrams with respect to all controllable parameters of the system, in which we find a stable homogeneous, phonon-collapsed, and roton unstable phases. 
We have shown the convergence of the condensate depletion at the threshold of the roton instability.
For tilted dipoles, we predict achievability of the threshold of the roton instability at the finite-size scales as well as the possibility of local density waves with controlled short-range order.
According to our estimations, the effects are achievable in experiments with ultracold atoms and polar molecules.

\section*{Acknowledgements}

We thank G.V. Shlyapnikov, E.A. Demler, U.R. Fischer, O.V. Lychkovskiy, and N.F. Stepanov for fruitful discussions of the results as well as
participants of the 23rd Laser Physics Workshop for useful comments.
This work is supported by the RFBR (Grants No. 14-02-00937 and No. 14-08-00606). 
A.K.F. and Y.E.S. acknowledge support from the Dynasty Foundation. 
Yu.E.L. is supported by the HSE Program of Basic Research.

\end{document}